\documentclass[a4paper, oneside, twocolumn, notitlepage, 10pt]{extarticle_ecoc}
\usepackage{ecoc}

\usepackage{romannum}

\def\ve#1{{\mathchoice{\mbox{\boldmath$\displaystyle #1$}}%
{\mbox{\boldmath$\textstyle #1$}}%
{\mbox{\boldmath$\scriptstyle #1$}}%
{\mbox{\boldmath$\scriptscriptstyle #1$}}}}
\def\j{\mathrm{j}}
\def\e{\mathrm{e}}

\newcommand{\norm}[1]{\left\lVert#1\right\rVert}

\newcounter{MYtempeqncnt}
\begin{document}
\selectlanguage{english}    % Standard Language

%-------------------------------------------------- Title -----------------------------------------------------%

\title{Perturbation-based Sequence Selection for Probabilistic Amplitude Shaping}%

%------------------------------------------------- Authors-----------------------------------------------------%

\author{
    Mohammad Taha Askari\textsuperscript{(1)} and Lutz Lampe\textsuperscript{(1)}
}

\maketitle                  % Create title and author

%------------------------------------------ Description of Authors ----------------------------------------------%

\begin{strip}
    \begin{author_descr}

        \textsuperscript{(1)} Department of Electrical and Computer Engineering, University of British Columbia, Vancouver, BC V6T 1Z4, Canada,
        \textcolor{blue}{\uline{mohammadtaha@ece.ubc.ca}} 

    \end{author_descr}
\end{strip}

% \setstretch{1.1}
%-------------------------------------------------- Footnote -------------------------------------------------------%
\renewcommand\footnotemark{}
\renewcommand\footnoterule{}
%\let\thefootnote\relax\footnotetext{text}

%-------------------------------------------------- Abstract ---------------------------------------------------------%

\begin{strip}
    \begin{ecoc_abstract}
        % NOTE: Don't use a blank line here but start abstract right away to avoid an extra line break
        We introduce a practical sign-dependent sequence selection metric for probabilistic amplitude shaping and propose a simple method to predict the gains in signal-to-noise ratio (SNR) for sequence selection. The proposed metric provides a $0.5$~dB SNR gain for single-polarized 256-QAM transmission over a long-haul fiber link. %, while it has lower complexity than the split-step Fourier method (SSFM).
        \textcopyright2024 The Author(s)
    \end{ecoc_abstract}
\end{strip}

%-------------------------------------------------- Introduction Section -------------------------------------------------------%

\section{Introduction}
Probabilistic amplitude shaping (PAS) has been adopted widely since its introduction in \cite{bocherer2015bandwidth}, due to its rate adaptability and systematic integration with forward error correction (FEC). Recently, PAS has also been utilized in conjunction with the optimization of the high dimensional distribution of shaped symbol sequences by a technique known as sequence selection. Originally, sequence selection was implemented based on rejection sampling, where a number of symbol sequences are generated %by a distribution matcher (DM) 
from an unbiased source\footnote{Examples are independent and identically distributed (i.i.d) symbols with Gaussian distribution or symbols generated from a distribution matcher (DM).}, and only a subset of those is retained based on a selection metric \cite{civelli2021sequence}. Using this, reference \cite{civelli2021sequence} and its extension \cite{secondini2022new} provided quantitative results for a new lower bound on the capacity of the optical fiber channel. However, the selection metric was more conceptual than practical since (i) it required the fiber channel simulation with the split-step Fourier method (SSFM) and (ii) did not provide a mapping between information bits and selected sequences.

%As the intention of the work was to provide a lower bound on the capacity of the optical fiber channel, the first attempt to use sequence selection for optical fiber channel, the authors in \cite{civelli2021sequence} provided a lower bound on the channel's capacity by using sequence selection with signal-dependent nonlinear interference (NLI) metric obtained after lossless split-step Fourier method (SSFM) simulation of the Manakov equation. Despite providing an interesting lower bound on the channel capacity, this work lacks practicality due to overly complex selection metric and lack of a constructive mapping between information bits and selected symbols. 

The first pragmatic implementation of sequence selection as part of a PAS scheme is list-encoding constant composition distribution matching (L-CCDM)\cite{wu2022list}. In L-CCDM, candidate sequences are generated by concatenating so-called flipping bits and information bits at the input of the CCDM, and selection is performed using the energy dispersion index (EDI) as a metric. The EDI is an energy metric % measure for the variance of windowed energy sequences, and it
that has been shown to be negatively correlated with the effective signal-to-noise ratio (SNR) of the nonlinear fiber channel for PAS with CCDM \cite{wu2021temporal}. Another energy metric for sequence selection, referred to as lowpass filtered symbol amplitude sequence (LSAS), was introduced in \cite{askari2022nonlinearity} and   \cite{askari2023probabilistic}. LSAS is based on a simplified first-order perturbative model and was shown to outperform EDI for sequence selection \cite{askari2023probabilistic}. % by considering inter-channel and inter-polarization effects.

The mentioned earlier studies on practical sequence selection demonstrated significant performance improvements in nonlinear optical fiber links. However, they had two shortcomings in common. First, a carrier phase recovery (CPR) was not considered at the receiver. Second, the EDI and LSAS metrics are purely based on the energy of symbols. The former has been discussed in \cite{civelli2023nonlinear, civelli2024sequence}. The authors highlighted that since a CPR partially compensates the nonlinearity induced phase noise, (i) any study on sequence selection for nonlinear interference noise (NLIN) mitigation needs to be verified with the presence of a practical CPR at the receiver, and (ii) a solely energy-based selection metric may fall short to provide substantive improvements in practical systems.
%thus, discounting the effect of shaping block length on fiber nonlinearity, while the NLI-based sequence selection gain remains intact. This finding suggests that any study on fiber nonlinearity, including sequence selection, needs to be verified with the presence of a practical CPR at the receiver. 
The work in \cite{civelli2024sequence} took one further step towards practical sequence selection by introducing several schemes for candidate generation. Of particular interest is a multi-block FEC architecture that makes use of the bootstrap scheme from \cite{bocherer:2011} to separate sign-dependent sequence selection and FEC. Since accounting for signs (or phases) in addition to selecting amplitudes (i.e., energies) of symbols has been shown to be almost as effective as also selecting signs \cite{civelli2024sequence}, this architecture reduces the effort for sequence generation to the necessary.  % by multi-block FEC architecture, while maintaining lower complexity compared to the flipping bits method. 

%This study also demonstrates that energy-based sequence selection is limited in the presence of a practical CPR and optimized shaping block length, highlighting the importance of the symbol's sign for sequence selection.

The recent investigation in \cite{liu2024sequence}  tackles the remaining problem of finding a selection metric that is sensitive to symbol signs but does not require an SSFM simulation. For this, the authors introduce the dispersion-aware EDI (D-EDI) as an extension of EDI. D-EDI attempts to account for the signs by averaging EDI metrics for several partially propagated sequences along a linear dispersive lossless fiber. %While D-EDI performs well in digital multi-band (DMB) transmission, it fails to account for inter-channel and inter-polarization nonlinear impairments, showing room for improvement. 
D-EDI inherits the problems of EDI though, which are rooted in the simple windowing of energy sequences and has been improved on by the LSAS metric \cite{askari2023probabilistic}. As pointed out in \cite{liu2024sequence}, it is mostly applicable for digital multicarrier transmission with low baud rate per subcarrier. 

In this paper, we propose a novel perturbation-based alternative to SSFM as a sequence selection metric. It consists of an energy-dependent NLIN term similar to LSAS and a symbol-dependent term that also accounts for the signs of symbols. As a second contribution, we present the use of the perturbative model in combination with the enhanced Gaussian noise (EGN) model \cite{carena2014egn} for the nonlinear fiber channel to estimate the effective SNR. This permits us to make predictions for the benefits of sequence selection and guide, for example, the choice of the number of sequences that should be generated. %inally, we show that sequence selection with this new metric, provides extra gain corresponding to the use of the symbol's sign, while it is capable of taking into account inter-polarization and inter-channel NLIs, making it an ideal candidate for DMT signalling where the symbols of all subchannels are available for sequence selection. 

\section{Sequence Selection Scheme}
\label{sec:seq_sel_scheme}
We apply the multi-block FEC-independent architecture from \cite{civelli2024sequence} that does not involve the FEC encoding in the sequence selection process. 

Let $K$ be the FEC blocklength in terms of quadrature amplitude modulation (QAM) symbols, and $L$ be the length used for sequence selection, where $m=K/L$ is assumed to be an integer. We use PAS with an ideal DM that realizes a  Maxwell Boltzmann (MB) distribution and produces two sequences of each $L$ amplitudes for the in-phase and quadrature-phase parts of $L$ QAM symbols. Because of the bootstrap FEC scheme, the signs for the in-phase and quadrature-phase components are known. We next generate $N$ QAM candidate sequences of length $L$ by symbol interleaving. %\footnote{We note that the Deinterleaving the selected sequence required $\log_2(N)$ bits, which FEC encoded and transmitted with the next FEC block.}.
% through the symbol interleaving method as in \cite{civelli2024sequence}.
This process is repeated for $m$ information blocks according to the FEC blocklength, so that we have $N$ candidate sequences for each of the $m$ information blocks. 

The $n$-th candidate sequences, $n=1,2,\ldots,N$, for the $m$ different information blocks are concatenated into one long sequence, which is input to the NLIN estimator using the first-order perturbative model as described in the next section. We use the $\ell_2$ norm of the estimated NLIN associated with each of the $N$ candidate sequences corresponding to the same information block to perform selection. Finally, the selected candidates are concatenated to form a sequence of $K$ QAM symbols used for transmission. 

We note that we use a block memoryless selection procedure, i.e., we ignore inter-block NLIN effects. Furthermore, side information of $\log_2(N)$ bits to identify the selected sequence at the receiver needs to be transmitted. While  \cite{civelli2024sequence} inserted pilot symbols for this, we suggest including the $\log_2(N)$ redundant bits into the input for the FEC encoding so that side information is FEC protected.

% For each block of information, we use an ideal distribution matcher with Maxwell Boltzmann (MB) distribution to generate $L$ amplitudes for the in-phase and quadrature-phase parts of QAM symbols, while employing the symbol interleaving method as in \cite{civelli2024sequence}, we end up with $N$ QAM candidates with length $L$. 
%We assume sign bits are known before sequence selection, so a single sequence of signs is used for all candidates, which is shown to be sufficient for sequence selection via the multi-block FEC independent scheme introduced in \cite{civelli2024sequence}. 
%Next, we concatenate the sequences generated for independent blocks of information and estimate the signal-dependent NLI using the first-order perturbative model as described in the next section. After NLI estimation, we use the $\mathrm{L}2$ norm of NLI to rank every $N$ sequence corresponding to the same block of information, and the selected candidates are concatenated and used for transmission. We note that we used a block memoryless selection procedure, where we do not consider the inter-block NLI effects.

 % shaping rate adjustment
 \begin{figure*}[!t]
%\normalsize
\setcounter{MYtempeqncnt}{\value{equation}}
 \setcounter{equation}{2}
% \vspace*{-2mm}
%%%%%%%%%%%%
% \begin{equation} \label{eq:add}
% \begin{split}
%  \Delta^{(c)}(n) = {\j}\gamma \sum_{m \in \mathbb{Z}\backslash\{0\}} \sum_{k \in \mathbb{Z}\backslash\{0\}} \bigg[ \bigg( s^{(c)}(n+m)[{s^{(c)}(n+m+k)}]^* s^{(c)}(n+k) \bigg)h^{(c,c)}(m,k) \\
%  + \!\!\!\sum_{c'\in {\cal C}\backslash\{c\}} \bigg(\!2s^{(c')}(n\!+\!m)[s^{(c')}(n\!+\!m\!+\!k)]^* s^{(c)}(n\!+\!k) \bigg)h^{(c,c')}(m,k) \bigg].
% \end{split}
% \end{equation}
%%%%%%%%%%%%
% \begin{equation} \label{eq:add}
% \begin{split}
%  \Delta^{(c)}(n) = {\j}\gamma \sum_{m \in \mathbb{Z}\backslash\{0\}} \sum_{k \in \mathbb{Z}\backslash\{0\}} \bigg[&\bigg( s^{(c)}(n+m)[{s^{(c)}(n+m+k)}]^* s^{(c)}(n+k) \bigg)h^{(c,c)}(m,k) \\
%  + & \sum_{c'\in {\cal C}\backslash\{c\}} \bigg(2s^{(c')}(n+m)[s^{(c')}(n+m+k)]^* s^{(c)}(n+k) \bigg)h^{(c,c')}(m,k) \bigg].
% \end{split}
% \end{equation}
%%%%%%%%%%%%
\begin{equation} \label{eq:add}
\begin{split}
 \Delta^{(c)}(n) = {\j}\gamma \sum_{c'\in {\cal C}} \sum_{m \in \mathbb{Z}\backslash\{0\}} \sum_{k \in \mathbb{Z}\backslash\{0\}} s^{(c')}(n+m)[s^{(c')}(n+m+k)]^* s^{(c)}(n+k) h^{(c,c')}(m,k).
\end{split}
\end{equation}
 \setcounter{equation}{\value{MYtempeqncnt}}
\vspace*{-4mm}
\hrulefill
\vspace*{2mm}
\end{figure*}

% \vspace{2mm}
\section{Perturbation-based Metric}
The proposed sequence-selection metric is based on the combination of the perturbation model in \cite{dar2014inter} and the additive-multiplicative model in \cite{liang2014multi} for the signal-signal interactions in optical fiber propagation. For concreteness, we consider a single-polarization wavelength division multiplexing (WDM) system and note that the extension to dual-polarization is straightforward \cite{askari2023probabilistic}. The symbol-rate sampled and dispersion compensated received signal in channel $c\in \cal C$ is written as
\begin{equation}
\label{eq:am_model}
r^{(c)}(n) = s^{(c)}(n) \e^{\j \theta^{(c)}(n)} + \Delta^{(c)}(n),
\end{equation}
where $s^{(c)}(n)$ is the transmitted symbol. %and received a sample at the discrete time $n$ and channel $c\!\!\in\!{\cal C}$, respectively. In this manuscript, we only derive the equations for the single-polarization system, while the extension to dual-polarization is straightforward \cite{askari2023probabilistic}. 
The NLIN in Eq.~\eqref{eq:am_model} consists of the multiplicative phase noise 
\begin{equation}
\label{eq:mult}
\theta^{(c)}(n) = \gamma \sum_{c' \in {\cal C}} \sum_{m \in \mathbb{Z}} |s^{(c')}(n+m)|^2 h^{(c,c')}(m,0),
\end{equation}
and the additive NLIN term shown in Eq.~\eqref{eq:add} at the top of this page. 
\setcounter{equation}{3} 
$\gamma$ is the fiber nonlinearity parameter, %$\j$ is the imaginary unit, 
and $h^{(c,c')}$ are the perturbation coefficients capturing the effect of channel $c'$ on channel $c$ \cite{askari2023probabilistic}.

For the purpose of sequence selection, we  %approximate the effect of CPR on nonlinear phase noise by 
 subtract the average phase rotation $\bar \theta^{(c)}$ to obtain the model % additive-multiplicative model after CPR as
\begin{equation}
\label{eq:am_model_cpr}
\hat{r}^{(c)}(n) = s^{(c)}(n) \e^{\j (\theta^{(c)}(n) - \bar \theta^{(c)})} + \Delta^{(c)}(n).
\end{equation}
%where $\bar \theta^{(c)}$ is the average phase noise over $n$. 
Finally, the selection metric for a sequence of symbols $\ve{s^{(c)}} = [s^{(c)}(0), \dots, s^{(c)}(L-1)]$ and the predicted received samples $\ve{\hat{r}^{(c)}} = [\hat{r}^{(c)}(0), \dots, \hat{r}^{(c)}(L-1)]$ %based on \eqref{eq:am_model_cpr} 
is the squared $\ell_2$ norm
\begin{equation}
\label{eq:am_metric}
\lambda_{\mathrm{AM}} = \norm{\ve{\hat{r}^{(c)}}-\ve{s^{(c)}}}^2.
\end{equation}
%where $\norm{.}$ denotes the $\mathrm{L}2$ norm operator. 

We refer to Eq.~\eqref{eq:am_metric} as the additive-multiplicative metric. It takes into account both energies and signs of symbols. When only considering the multiplicative NLIN part in Eq.~\eqref{eq:am_model_cpr}, we obtain the energy-based LSAS metric from  \cite{askari2023probabilistic} as a special case. We remark that while the multiplicative and additive terms in Eq.~\eqref{eq:mult} and Eq.~\eqref{eq:add}  are described for the general case of WDM channels in set $\cal C$, we only consider the intra-channel NLIN for sequence selection in this work. However, extension to inter-channel NLIN, which is relevant for digital multicarrier systems, is straightforward using Algorithm~1 from \cite{askari2023probabilistic}. 

\section{Performance Prediction}
 We suggest also applying the additive-multiplicative model for NLIN in Eq.~\eqref{eq:am_model} for predicting the gains achievable with sequence selection. For this, we consider the EGN model and relate the NLIN power $P_{\mathrm{NLIN}}$ to the transmit power $P$ as
\begin{equation}
% \vspace{-1mm}
\label{eq:egnnlin}
P_{\mathrm{NLIN}} = \eta P^3,
\end{equation}
 where  the nonlinearity coefficient $\eta$ depends on signal-signal interactions and modulation format \cite{dar2014inter}. Considering two sequence-selection shaping schemes corresponding to coefficients $\eta_1$ and $\eta_2$, the ratio of their effective SNRs at optimum transmit power is given by
 \begin{equation}
\label{eq:egnsnr}
\frac{\mathrm{SNR}_1}{\mathrm{SNR}_2} = \left(\frac{\eta_2}{\eta_1}\right)^{1/3}.
\end{equation}
 From Eq.~\eqref{eq:egnnlin} and Eq.~\eqref{eq:egnsnr}, measuring the $P_{\mathrm{NLIN}}$ using Eq.~\eqref{eq:am_model} permits us to predict the SNR gain in dB as 
  \begin{equation}
 \label{eq:egngain}
 \mathrm{SNR}_{1}- \mathrm{SNR}_{2} = \frac{1}{3}\left(P_{\mathrm{NLIN},2}-P_{\mathrm{NLIN},1}\right)\;\mbox{dB}.
 % \vspace{-1mm}
 \end{equation}
% \begin{equation}
%\label{eq:egngain}
%\mathrm{SNR}_{1}- \mathrm{SNR}_{2} = %\frac{1}{3}\left(\eta_2-\eta_1\right).
%\vspace{-1mm}
%\end{equation}
% of a system with amplified spontaneous emission (ASE) noise and signal-dependent impairments as
%\begin{equation}
%\label{eq:egn}
%\mathrm{SNR} = \frac{P}{P_{\mathrm{ASE}} + \eta P^3},
%\end{equation}
%where $P_{\mathrm{ASE}}$ is the ASE variance, $P$ is the launch power, and $\eta$ is a proportionality factor, which is dependent on signal-signal interactions and modulation format \cite{dar2014inter}. We note that the variance of NLI predicted by the additive-multiplicative model estimates $\eta P^3$, which results in the SNR at the optimal launch power as
%\begin{equation}
%\label{eq:egn_opt}
%\mathrm{SNR_{opt}} = \bigg(\frac{P_{\mathrm{ASE}}}{2\eta}\bigg)^{\frac{1}{3}}.
%\end{equation}

In the next section, we use the metric in Eq.~\eqref{eq:am_metric} for sequence selection, and we will compare the SNR gain prediction using Eq.~\eqref{eq:egngain} with the measured SNR gain by full SSFM simulation.

\section{Numerical Results}
The system setup is adopted from \cite{askari2023probabilistic, wu2022list}. 
We use the SSFM to simulate communication over $3$ single-polarized WDM channels, each operating at $32$~GBd baud rate, and with a channel spacing of $50$~GHz. The sender applies PAS with MB shaping for 256-QAM and a rate of $2.5$~bits/amplitude and sequence selection with sequence length $L=256$. The link consists of $20$ spans of $80$~km of a standard single-mode fiber, with chromatic dispersion coefficient (CD) $17$~ps/nm/km, nonlinearity parameter $\gamma\!=\!1.37$~W$^{-1}$km$^{-1}$, and fiber loss $0.2$~dB/km. An erbium-doped fiber amplifier (EDFA) with noise figure~$6$~dB is placed at the end of each span. At the receiver, CD is electronically compensated and a linear pilot-aided CPR, adopted from \cite{neshaastegaran2019log}, with a $1\%$ pilot rate is applied. We estimate the achievable information rate (AIR) using the mismatched decoding method described in \cite{alvarado2017achievable}, and deduct for the selection rate loss of $\log_2(N)/L$~bits/2D. Each simulation result is based on $2^{18}$ generated $256$-QAM symbols, and the performance results for the WDM center channel are considered.

\begin{figure}[t!]
    \centering
    \includegraphics[width=\linewidth]{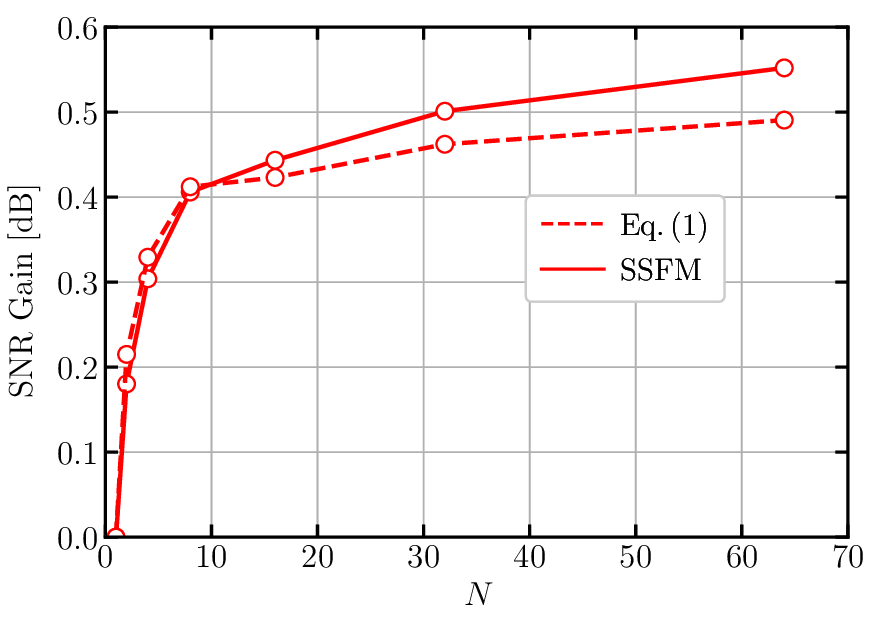}
    \caption{SNR gain for sequence selection over $N$ candidates.}
    \label{fig:snr}
\end{figure}

Fig.~\ref{fig:snr} shows the SNR gain at optimal launch power achieved by sequence selection using the additive-multiplicative metric in Eq.~\eqref{eq:am_metric} with $N \in \{1, 2, 4, \dots, 64\}$ candidates, where $N\!=\!1$ corresponds to the system without sequence selection. We observe that sequence selection can provide more than $0.5$~dB gain in SNR. Most of the gain is achieved using only $N\!=\!8$ candidates, which corresponds to allocating only $3$ out of $2048$ bits for selection of a sequence. In addition to the SNR gain measured after SSFM simulation, we used Eq.~\eqref{eq:am_model} to approximate a noiseless fiber channel, and after applying CPR at the receiver, we used Eq.~\eqref{eq:egngain} to estimate the SNR gains. We observe that SNR measurements and predictions are well aligned in Fig~\ref{fig:snr}. This supports the use of the additive-multiplicative model both for sequence selection and for performance prediction. % shows that the gains predicted by the additive-multiplicative model is well aligned with the measured SSFM gains, which again, confirms the appropriateness of NLI predicted by the additive-multiplicative metric for sequence selection.

\begin{figure}[t!]
    \centering
    \includegraphics[width=\linewidth]{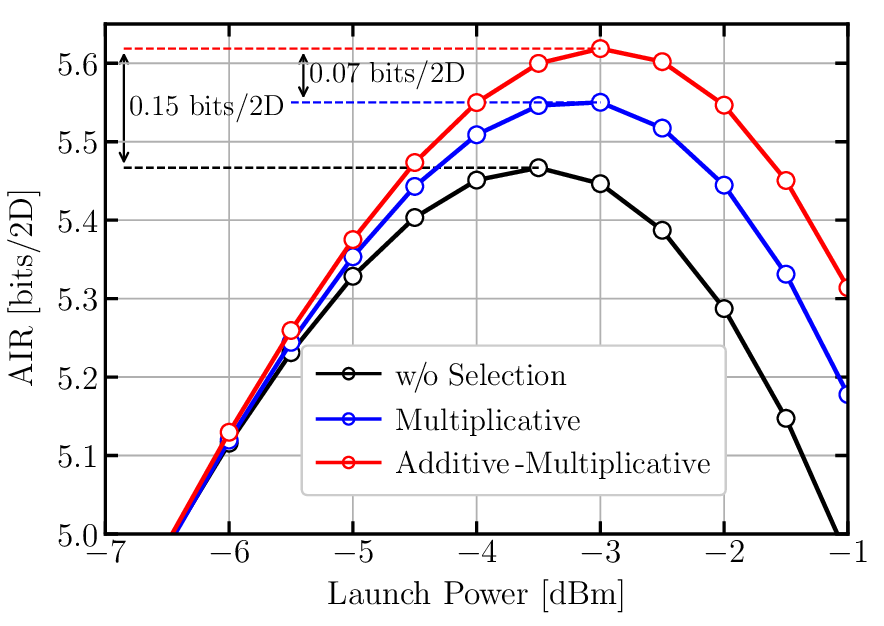}
    \caption{AIR for %vs.\ launch power. Sequence 
    sequence selection with $N=64$ candidates and different selection metrics.}
    \label{fig:air}
\end{figure}

Fig.~\ref{fig:air} shows the AIR versus launch power for the shaping without selection ($N\!=\!1$), selection with $N\!=\!64$ candidates using (i) only the multiplicative term and (ii) both additive and multiplicative terms of Eq.~\eqref{eq:am_model_cpr}. We observe that all three schemes perform the same in the linear regime, while sequence selection provides clear benefits as power increases. Moreover, sequence selection using the additive-multiplicative metric provides $0.07$~bits/2D extra gain compared to the energy-based metric. This result shows that the gains observed in \cite{civelli2024sequence} due to considering sign bits in sequence selection are realized using the practical metric proposed in this work.

\section{Conclusions}
We introduced a sequence selection metric based on the first-order perturbative model, which takes into account the signs and energies of shaped symbols. Simulation results show a $0.15$~bits/2D AIR improvement by sequence selection using this metric compared to PAS without selection. The proposed metric can naturally account for inter-polarization and inter-channel NLIN effects. Finally, we suggest that the underlying additive-multiplicative NLIN model is effective for SNR performance prediction. 
%, while maintaining a lower complexity compared to SSFM, which makes it an ideal candidate for practical sequence selection. 

%-------------------------------------------------- Acknowledgements Section -------------------------------------------------------%
\clearpage
%\section{Acknowledgements}
%This work was supported by Huawei Tech., Canada, and enabled in part through support provided by the Digital Research Alliance of Canada (www.alliancecan.ca).
%-------------------------------------------------- Bibliography Section -------------------------------------------------------%
% see also https://tex.stackexchange.com/questions/55030/text-before-references-but-after-bibliography-title-with-bibtex as of 2024-02-29
% \defbibnote{myprenote}{%
% Citations must be easy and quick to find. More precisely:
% \begin{itemize}
%     \item Please list all the authors. 
%     \item The title must be given in full length. 
%     \item Journal and conference names should not be abbreviated but rather given in full length.
%     \item The DOI number should be added incl. a link.
% \end{itemize}
% }
\printbibliography

\vspace{-4mm}

%%%%%%%%%%%%%%%%%%%%%%%%%%%%%%%%%%%%%%%%%%%%%
%---------------------------------------------- End of Document -----------------------------------------------%
\end{document}